\documentstyle[11pt,newpasp,twoside,epsf]{article}
\begin{document} \title{The McKee/Ostriker Model: Paradigm?}  

\title{ Presented at The Fourth Tetons Summer Conference:
Galactic Structure, Stars, and the Interstellar Medium}

\author{Carl Heiles} \affil{Astronomy Department, University of
California, Berkeley, CA 94720-3411}

\begin{abstract} We summarize the current observational status of the
interstellar medium in terms of the McKee/Ostriker model. We find both
agreement and disagreement---hence the title of the paper. Some of the
newest observations differ from older ones,  which underscores the need
for both more and better data.  \end{abstract}


\section{Introduction}

          In 1977 Mckee \& Ostriker (MO) extended the two-phase model of
Field, Goldsmith, \& Habing (1969; FGH) by including supernova and wrote
their famous paper entitled ``A theory of the interstellar  medium -
Three components regulated by supernova explosions in an inhomogeneous
substrate''.   Their ISM is dominated by individual supernova (SN)
explosions; today some believe we should replace the individual
explosions by correlated ones producing superbubbles.  The Hot Ionized
Medium (HIM) fills the interior of SN remnants and powers their blast
waves, which sweep up the gas and magnetic field inside the bubble and
pile them into the shell.  Soon this shocked gas starts to cool and
recombine rapidly, forming the Cold Neutral Medium (CNM) with an
amplified field.  Soft X-rays produced by immediately adjacent HIM
penetrate the outsides of CNM clouds, heating the gas to form the
partially ionized Warm Neutral Medium (WNM), whose more partially
ionized portions are the Warm Ionized Medium (WIM). Their model is
impressive in its physical development and internal consistency.  It
makes many predictions and stands as the accepted paradigm against which
most discussions and interpretations are compared. 

	The FGH and MO models rely on a broad range of physical
processes, ranging from atomic interactions with energetic particles to
gasdynamics, and the uninitiated may be easily overwhelmed. Fortunately,
there is a straightforward path to nirvana, namely excellent, classical
reviews that remain useful for experts. Dalgarno \& McCray (1972)
present the physics of heating and cooling for atomic regions, the basis
for the FGH model. MO's paper is a bit intimidating  for many of us, but
their Figures $1 \rightarrow 3$ and Table 1 present an excellent
three-minute overview of the basic picture. McCray \& Snow (1979)
present a highly readable discussion of the details and basic physics
behind all aspects of the MO model, together with some critical comments
that remain relevant today.  At the first Tetons meeting, Kulkarni \&
Heiles (1987; KH) interpreted ISM observations of the cool and warm
phases in terms of the MO paradigm with great success. Cox (1995)
presents a historical overview of our ISM understanding, and lack of it,
and emphasizes some specific problems with the MO model.

          The MO model doesn't cover everything. The magnetic field and
cosmic rays play important roles in the overall dynamics of the ISM. 
Their pressures are larger than the typically adopted thermal pressures
of the gas components.  Even for the large-scale $z$-distribution, where
the pressure gradients may be small, their pressures contribute to the
hydrostatic balance (Boulares \& Cox 1990).  Also, the magnetic field
links the different gas phases with each other, and also with the cosmic
rays, so that components cannot act independently.  Rather, analysis of
the dynamics and equilibria must include the whole ensemble, and this
leads to new modes of behavior such as magnetically-linked clouds
(Elmegreen 1994) and the Parker (1966) instability.  As important as
these two nonthermal, non-material components are, we will not emphasize
them in this review. 

	Rather, we will define and review {\it five} (not MO's three)
diffuse gas phases. These are the HIM, the HII-region like Reynolds WIM
(RWIM), the MO-like WIM (MOWIM), the WNM, and the CNM. We begin with a
definition of and brief introduction to the five phases in \S2 and then
go on to discuss various aspects in more detail.

\section{Introduction to the Five ISM Phases \label{intro-to-four}}

          Table~\ref{bigtable} presents the essential properties of the
five diffuse gas phases.  The tabular form inspires confidence, but this
is misleading because there is uncertainty and disagreement; the table
reflects my personal bias.  Perhaps most blatant, the idea that the
MOWIM is an observationally important phase that contributes most of the
total electron column is not (yet?) widely accepted. 

{\tiny
{\begin{table}\tiny} 
\begin{center}
\caption{  THE DIFFUSE ISM PHASES \label{bigtable} }
\begin{tabular}{cllll}
\\
Quantity                    & HIM                           & RWIM                         &  WNM/MOWIM                                      & CNM  \\ 
\tableline
h
$T$, K                      &  ?$\rightarrow 10^7$          &  $\sim 8000$                 & $500 \rightarrow 8000$\tablenotemark{A}         &  $10 \rightarrow 75$\tablenotemark{A}   \\
${P \over k}$, cm$^{-3}$ K  &  $\ga 20000$\tablenotemark{B} & $\ga 3400$\tablenotemark{B}  & $200? \rightarrow 4000$\tablenotemark{A}      &  $1500 \rightarrow 10000$              \\
$n$, cm$^{-3}$              &  0.003                        &  $\sim 0.08$                 & $0.1 \rightarrow 0.4$\tablenotemark{A}          & $ 20 \rightarrow 250$               \\
$X_e$                       & 1                             & 1                            & $\sim 10^{-2} \rightarrow 0.5$\tablenotemark{A} & $\sim 2 \times 10^{-4} \ (= {{\rm C} \over {\rm H}})$  \\
$f$                         & 0.5\tablenotemark{C}          & 0.1\tablenotemark{C}         & 0.5\tablenotemark{C}                            & 0.01\tablenotemark{C}   \\
$N_{20 \perp}$              & $\sim 0.3$?\tablenotemark{E}  & 0.1?\tablenotemark{E}        &H, 1.9; e, 0.3?\tablenotemark{E}                   & 1.8                     \\
Heating                     & shocks                        & H photoion + ?               & grains, etc\tablenotemark{D}                    & C ionization, etc\tablenotemark{D}  \\
Ionization                  & el coll                       & H photoion                   & H XR, some CR                                   & C photoion, some CR       \\
Observing                   & XR, UVa, UVe                  & DM, EM, SM, UVa, FS          & HIem, UVa,IRe, FS                               & HIe, HIa, UVa, IRe, FS  \\ 
\end{tabular}
\end{center}

\tablenotetext{}{In column 1, $X_e$ is the ionization fraction ${n_e
\over n_e + n_{HI}}$; $f$ is volume filling fraction; $N_{20 \perp}$ is
the H or electron column density projected towards the Galactic pole  in
units of $10^{20}$ cm$^{-3}$. In other columns, XR is X-Ray emission; DM
pulsar dispersion; EM H$\alpha$ emission; SM pulsar scattering; FS fine
structure lines in emission; IRe continuum IR emission from grains; UVa
optical/UV absorption lines against stars; UVe UV emission lines; HIe
and HIa 21-cm line emission and absorption.}

\tablenotetext{A}{ This quantity is critically discussed herein and 
these values may be either not correct or not generally accepted.}

\tablenotetext{B}{ These are typical pressures; wide fluctuations exist.}

\tablenotetext{C}{ Volume filling factors depend on $z$ and are highly 
uncertain; these values are for $z=0$.}

\tablenotetext{D}{ In contrast to the other phases, multiple 
heating mechanisms are important for the WNM and CNM; see Figure 3
of Wolfire et al (1995a).}

\tablenotetext{E}{ The $N_e\perp$ for the HIM is a theoretical value
(Wolfire et al 1995b). The relative contributions of the RWIM and the MOWIM 
to $N_{e,\perp}$ are abritrarily chosen and are highly uncertain
(\S5).}

{\end{table}}
}
 
\subsection{Introduction to the HIM}

	The HIM is produced by SN shocks and resides in SN remnants and
superbubbles. It can be mapped with its X-ray emission.  Cooler remnants
can be observed with 0.25 keV X rays if the intervening HI column
density is low enough ($N_{HI,20} \la 0.6$,  where $N_{HI,20}$ is the HI
column density in units of $10^{20}$ cm$^{-2}$).  This is tiny, so even
the thinnest cloud obscures the 0.25 keV emission.  Higher energies are
easier because the HI opacity is much smaller; for example, for $T \ga 2
\times 10^6$ K the 0.75 keV emission can be seen through $N_{HI,20} \la
20$.  Even this is serious for locating distant objects because this
column of HI is normally accumulated over a path length $\sim 700$ pc,
so we can only map nearby structures, of which there are two prominent
ones: the North Polar Spur and the Eridanus superbubble. The ROSAT
satellite's all-sky X-ray maps for 0.25, 0.75, and 1.5 keV (Snowden et
al 1997) are the best available.

	Gas having $T \la 0.7 \times 10^6$ K must exist, because the HIM
cools down as it ages.  But such cool gas cannot be seen with X rays. 
UV absorption lines of the He$^+$-like ions OVI, NVI, and CIV trace HIM
at temperatures $\sim 3 \times 10^5$ K and below.  Such lines provide
velocity resolution and column density, which greatly helps the
interpretation (e.g.  Shelton \& Cox 1994). 

	These lines can also be seen---weakly---in emission, which
provides the emission measure instead of column density, so a comparison
of absorption and emission provides the electron density $n_e$.  OVI
emission reveals gas at $T \sim 3 \times 10^5$ K and implies high
pressures, ${P \over k} \sim 40000$ cm$^{-3}$ K (Dixon, Davidsen, \&
Ferguson 1996) while CIV reveals $T \sim 10^5$ K and implies much lower
pressures, ${P \over k} \sim 2000$ cm$^{-3}$ K (Martin \& Bowyer 1990). 
Both datasets are noisy and sample only a few sightlines, so these
results should be regarded as provisional. 

\subsection{Introduction to the Reynolds WIM (RWIM)}

        The RWIM is widely distributed, HII-like highly ionized gas
having $n_e \sim 0.08$ cm$^{-3}$ and $T \sim 8000$ K (Haffner, Reynolds,
\& Tufte 1999).   The RWIM is often called the ``Reynolds layer'', after
the person who has pursued its study most vigorously; the culmination is
the WHAM survey, which provides the Galactic H$\alpha$ data cube for
most of the sky (Haffner, this meeting).   Its ionization requires a
significant continuing energy input; for conventional sources only
starlight is sufficient (Reynolds 1984), and despite concerns about how
the starlight penetrates neutral regions, it seems theoretically
possible and the RWIM can be described as HII region envelopes (Miller
\& Cox 1993; Dove \& Shull 1994; McKee \& Williams 1997; Anantharamaiah
1985, 1986).  In \S3.1 we extrapolate from the Eridanus superbubble and
argue that much of the RWIM is formed by ionization of supershell walls
by stellar UV photons from the OB clusters inside. The question mark
under RWIM heating in Table~\ref{bigtable} refers to a possible extra
heating component required to explain observed RWIM temperatures at high
$|z|$ (Reynolds, Haffner, \& Tufte 1999). 

	Taylor \& Cordes (1993; TC) use pulsar DM's to derive the
standard model of electron distribution in the Galaxy.  They use the
implicit assumption that pulsar DM's are produced by the RWIM; in
contrast, we argue below that the DM's are produced mainly by the MOWIM
(see below). The RWIM can also be observed with IR fine structure
emission lines (e.g.~Heiles 1994). 

\subsection{Introduction to the MOWIM}

	MO predicted this phase, considering it simply as partially
ionized WNM.  However, if it really produces most of the observationally
important pulsar DM's, then we believe its status should be elevated and
considered as a separate phase.   In \S4 and 5 we discuss H$\alpha$,
pulsar, and optical/UV absorption line data in some detail and argue
that, indeed, a significant fraction of the pulsar DM-producing
electrons does reside in the MOWNM. The idea that the WNM contributes
significantly to pulsar dispersion is contrary to current observational
thought---but in agreement with MO. 

\subsection{Introduction to the WNM and CNM}

	In the MO model, the CNM resides in the walls of SN remnant
walls. Soft X-rays produced by the adjacent HIM penetrate the outside
layer of CNM clouds, heating and partially ionizing the gas to form the
WNM; in our parlance, this partially-ionized WNM is the MOWIM.  Thus
the swept-up gas consists of two distinct neutral phases, each in
thermally stable equilibrium and with thermal pressures at least roughly
equal.  The shells can break up, forming clouds, and these in turn can
be overcome by a different passing SN shock, placing them inside either
a SN remnant or a superbubble wall.

        The CNM and WNM temperatures are derived by calculating the
equilibrium temperature as a function of thermal pressure.  With this
formulation and realistic heating and cooling functions there exist two 
stable ranges of equilibrium, the CNM and the WNM with temperatures 
$\sim 50$ and $\sim 8000$ K, separated by a region of unstable
temperatures.  {\it Stable} thermal pressure equilibrium, and thermal
pressure equality between physically adjacent phases,  is a cornerstone
of the MO theory, although the pressure varies widely from one region to
another, as observed (Jenkins, Jura, \& Lowenstein 1983). In \S6.3, we
find that a significant population of WNM lies at thermally {\it
unstable} temperatures---contrary to MO.  

	The 21-cm line in emission maps the total HI column density,
both the WNM and the CNM, in the usual case when the opacity is not too
high.  The latest survey of the northern sky is Hartmann \& Burton
(1997), with 36 arcmin resolution.  The opacity $\propto {1 \over T}$,
and the large temperature difference between WNM and CNM means that, in
effect, only the CNM is seen in absorption spectra against background
continuum sources.  We defer discussing 21-cm line and optical/UV
absorption line data to \S6. 

	IR emission from grains is a reasonably accurate tracer of
total column density of gas, including not only the atomic components
but also the ionized and molecular components; the most useful is the
100 $\mu$m IRAS maps (e.g. Schlegel, Finkbeiner, \& Davis 1998). 

	The physically most interesting tracer is the 158 $\mu$m fine
structure line of C$^+$, because this is the principal coolant.  Maps of
this line, with sufficient sensitivity and velocity resolution to
distinguish WNM from CNM, would be of great value in developing our
understanding of the physical processes in these phases.  The only
current data on local gas are Matsuhara et al (1997), and even though
they have insufficient velocity resolution, they are valuable in
providing accurate cooling rates.  Nakagawa et al (1998) provide a
large-scale survey of the interior Galactic plane. 

\section{The ISM and Superbubbles \label{superbubble-dominated}}

\subsection{The Eridanus Superbubble \label{eridanus-superbubble}}  

          The Eridanus superbubble has been studied previously by
several groups; the description below is a brief version of Heiles,
Haffner, \& Reynolds (1999) and is an extract of a forthcoming paper
which synthesizes multiwavelength data in an attempt to provide a more
complete physical description. 

\begin{figure} 
\plottwo{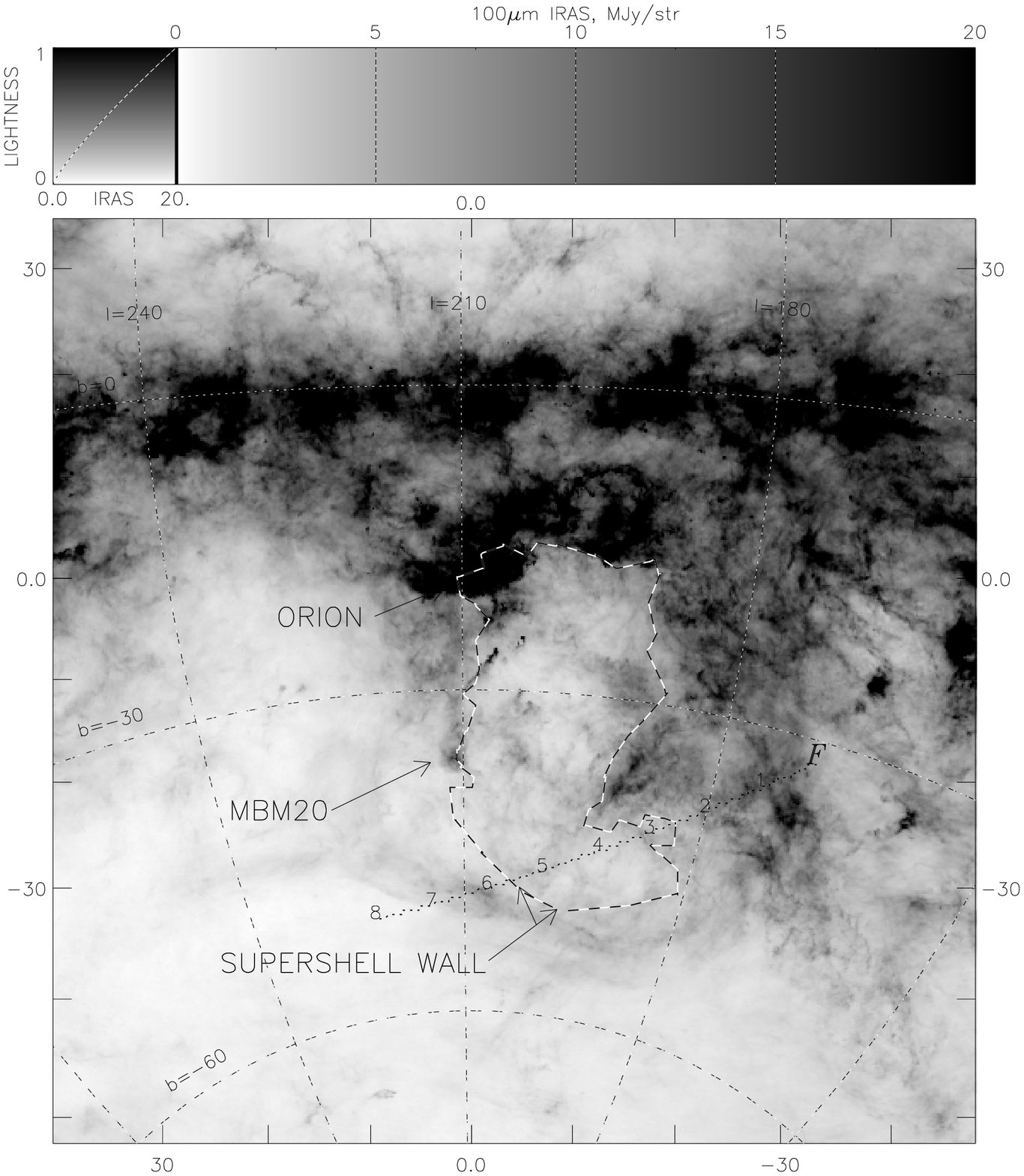} {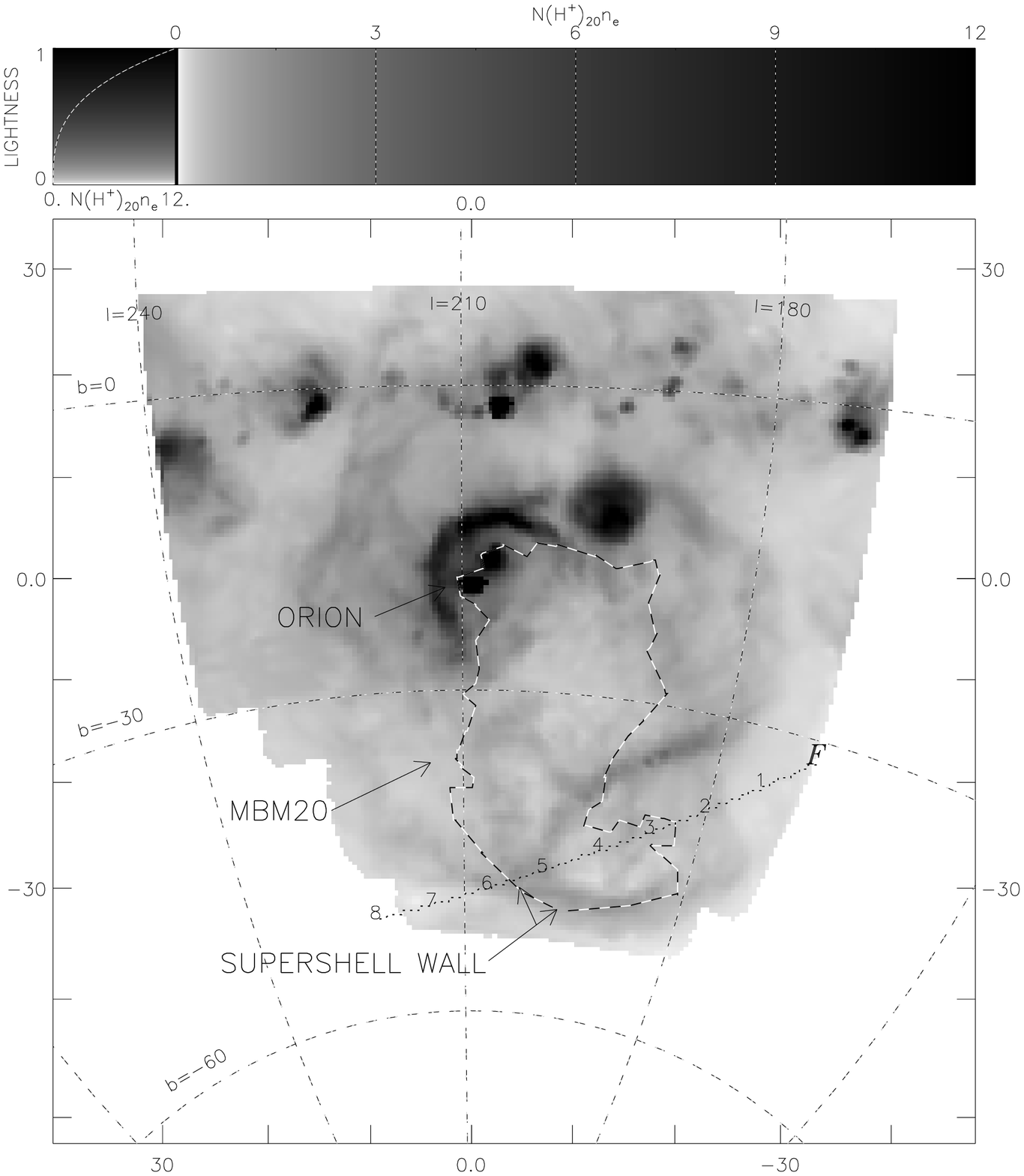} 

\caption{The Eridanus regions mapped in {\it (left)}  100 $\mu$m
emission from IRAS, which traces mainly atomic and molecular hydrogen,
and {\it (right)} H$\alpha$ emission from the WHAM survey.  The heavy
dashed line labeled ``SUPERSHELL WALL'' outlines the boundary of the
0.75 keV emission, which is very sharp.  The grey scales are calibrated
by the colorbars on top right of each panel; saturated parts of the
images exceed the colorbar maxima.  Galactic coordinates are labeled on
the dashed grid lines.  Larger versions are in Heiles et al (1999).
\label{fig-iras}
} 

\end{figure}

 	The Eridanus superbubble is a huge structure that was produced
by the winds and SN of previous generations of Orion stars.  The central
volume of the superbubble is full of very hot gas whose spatial
structure is well delineated by 0.75 keV X-ray emission, which has a
sharp, well-defined edge; we represent the edge with a dashed line on
figures that show other gas components.  The distance $\sim 500$ pc and
the overall size $\sim 30^\circ$ or 250 pc.  The 0.75 keV spatial
structure is duplicated by higher-energy very dim 1.5 keV emission.  The
presence of even a little of this higher energy emission implies a
temperature of several million degrees or more, hotter than
currently-accepted estimates. 

	In contrast, the lower-energy 0.25 keV emission (Figure 2, {\it
left}) is brightest near the higher-$|b|$ portions of the superbubble,
and also outside its boundary.  This outside emission is traditionally
thought to come from gas in the Galactic halo.  However, we believe that
this gas has leaked out of the rear wall of the superbubble and, using
21-cm and H$\alpha$-line data, have even been able to locate the
specific place where this has happened. 

          Figure~\ref{fig-iras} maps the diffuse 100 $\mu$m IRAS
emission, which traces mainly atomic and molecular hydrogen, and the
H$\alpha$ from the WHAM survey (Haffner, this meeting).  All gas phases
are prominent in the Eridanus Superbubble: the HIM, which emits the XM
emission; the atomic and molecular gas, which sits in the superbubble
wall; and the RWIM, which also sits in the superbubble wall. 

\bigskip

\noindent{\it The Onionskin structure.} The lower edge of the
superbubble wall, near $b \sim -50^\circ$, has a well-delineated
onionskin-type structure, which can be seen by comparing the two panels
in Figure 1: the inside layer of the wall is ionized and the outside
layer is neutral.  This structure is consistent with photoionization
caused by the hot Orion stars, because there is little or no neutral gas
inside of the bubble and photons stream, unimpeded, from the stars to
this high-$|b|$ gas. 

	Near the middle of the superbubble the photoionization structure
takes on a different character.  Instead of an onionskin-type structure
the walls are either mostly completely ionized or completely neutral. 
Specifically, over much of the more distant wall of the superbubble the
gas is mainly atomic, while the near wall is mainly ionized.  This is
consistent with the geometry of the OMC region where the dense molecular
clouds lie behind the hot stars, so the molecular clouds shield the far
wall of the superbubble from the ionizing stellar photons. 

	In Eridanus, the relationship between the neutral gas and the
RWIM is clear.  The WIM exists where ionizing photons are able to reach
the neutral wall.  Indeed, it is difficult to imagine any other source
of ionization or energy for the RWIM, so it is natural to extend this
concept to all of the WIM: namely, that the RWIM exists in regions that
would be occupied by neutral gas were it not for the accidental presence
of ionizing photons.  These photons ionize and heat the gas, raising its
pressure by at least a factor of two. 

\bigskip 

\noindent{\it The electron density and pressure.} Heiles et al
(1999) derive electron volume densities by comparing the H$\alpha$,
21-cm line, and 100 $\mu$m intensities.  They find typically $n_e \sim
0.8$ cm$^{-3}$, which corresponds to ${P \over k} = 2 n_e T \sim 12000$
cm$^{-3}$ K.  This is higher than the pressure of typical CNM clouds,
which makes sense because the wall is adjacent to an overpressured HIM
region. 

	We know of one other density determination, namely in the large
filament mapped by Haffner, Reynolds, \& Tufte (1998).  They obtain
$n_e \sim 0.18$ cm$^{-3}$, which corresponds to ${P \over k} \sim 2500$
cm$^{-3}$ K. 

\subsection{Other prominent nearby superbubbles}

\begin{figure} 
\plottwo{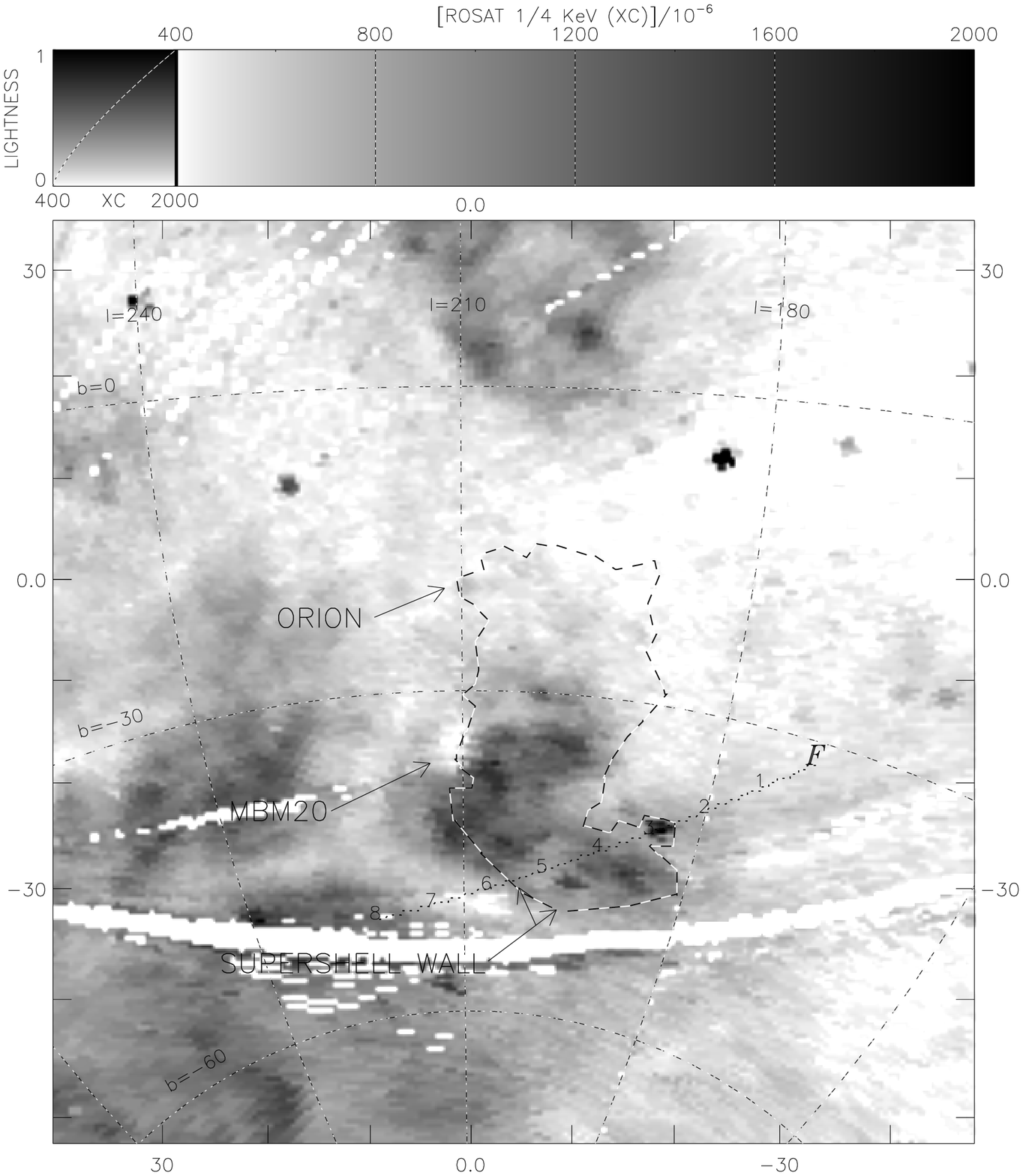} {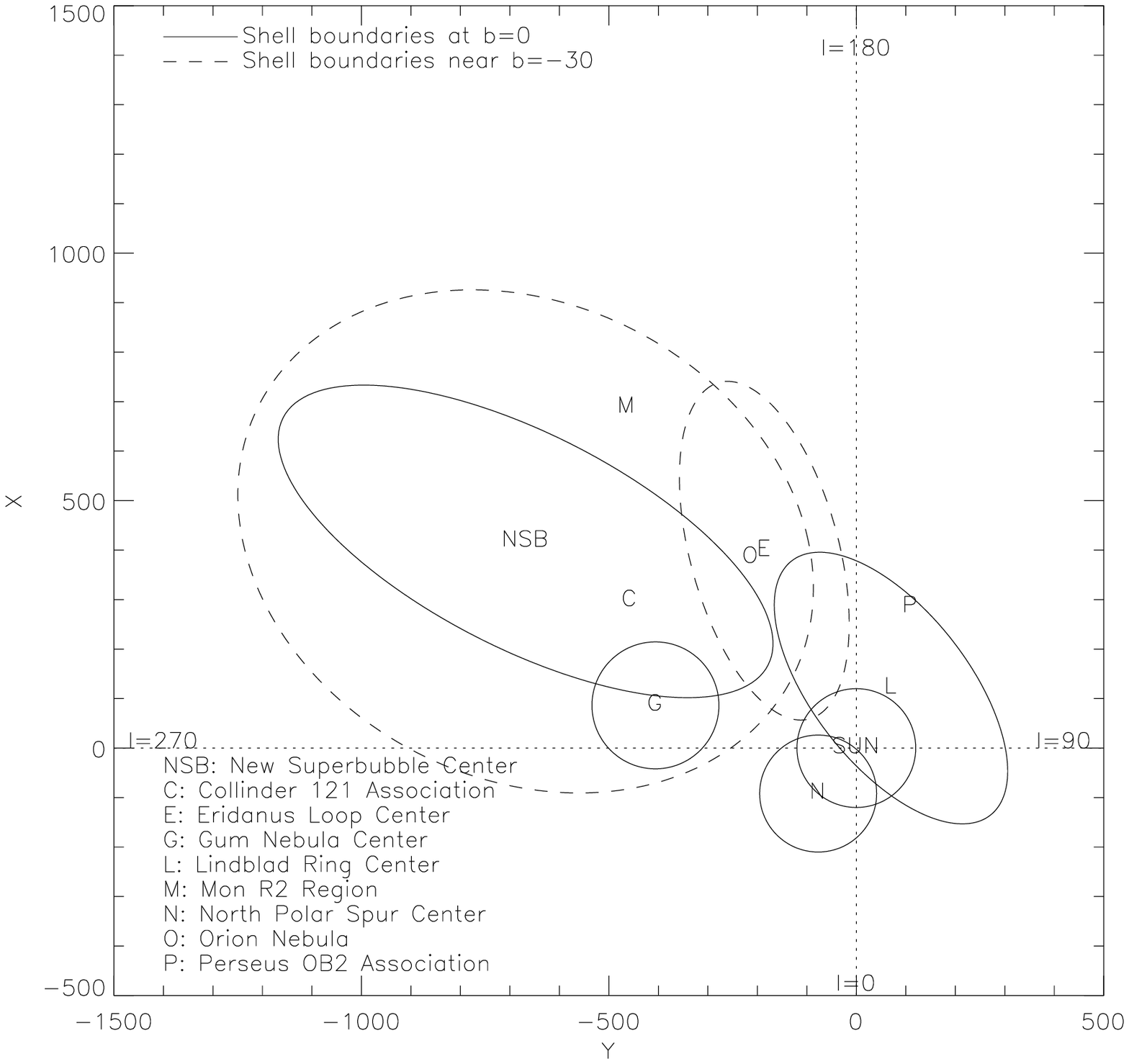}  

\caption{ {\it Left:} The Eridanus
region in 0.25 keV X-rays; for details, see the caption for Figure 1.
{\it Right:} Nearby HI shells and bubbles projected onto the Galactic
plane, viewed from the north Galactic pole (Heiles 1998).  Solid
circles or ellipses represent shells near $b=0^\circ$; dashed ones
represent shells near $b \sim -30^\circ$.  \label{heilesmap}}

\end{figure}

	We are fortunate in having the textbook example of Eridanus so
close and easy to study.  However, the probability of this fortunate
occurrence is not all that small: the Solar neighborhood is riddled with
superbubbles.  Another prominent one is the North Polar Spur superbubble
(e.g.~Egger 1998), which was produced by stars in the Scorpius/Ophiuchus
cluster and lies opposite the Eridanus superbubble in the sky.  Another
is the much older Local Bubble, in which the Sun is immersed.  Another
is the ``New Superbubble'' (Heiles 1998), a gigantic old structure
centered near $\ell \sim 230^\circ$, which may have been responsible for
inducing star formation in the Orion and the Gum regions.  And there are
numerous others that are older and consequently not very prominent, but
whose effects are rather clearly documented in the dynamics of the local
gas and in the location of related stellar associations (Bally, this
meeting).  

	RWIM filaments lie on the surface of the Eridanus superbubble. 
We believe that this also happens for the huge, prominent Haffner et al
(1998) RWIM filament, but don't have the space to present this argument.
Given these specific examples, we extrapolate and conclude that all
prominent RWIM filaments lie on the surfaces of superbubbles in regions
where ionizing photons happen to penetrate.  Reynolds (1984) also sees a
less structured H$\alpha$ emission component, which corresponds to a
more pervasively distributed WIM; this is probably the agglomeration of
older versions of the prominent filaments, HII region envelopes, and the
MOWIM (partially ionized WNM; \S4.1 and 5). 

\subsection{A superbubble-dominated ISM?}

          Heiles (1998) provides a map of the local superbubbles, which
we reproduce in Figure~\ref{heilesmap} {\it right}.  They are so
numerous that they run into one another and occupy common volumes.  This
suggests that their HIM's are interconnected, which would mean that the
local ISM is superbubble dominated. 

	However, previous studies find that superbubbles do not occupy a
large fractional ISM volume. Observationally, Heiles (1980) uses mainly
northern-hemisphere HI data to conclude that the superbubble volume
filling fraction is $10 \rightarrow 20\%$.  Theoretically, McKee (1993)
uses the luminosity distribution of star clusters to calculate the
effects of clustered SN and find that the superbubble filling fraction
should be $\sim 10\%$.  

	We can reconcile these seemingly inconsistent conclusions, but
not in a way that leads to a definitely preferred choice. If one wants
to minimize superbubble dominance, then one argues that some of the
objects on Figure~\ref{heilesmap} are detectable only by fairly weak
kinematic signatures, and that they are so old that  their interiors  no
longer contain HIM; also that some are identified only by
overinterpretation of the data, and not real. 

	In contrast, if one wishes to maximize superbubble dominance,
then one argues as follows: Observationally, that Heiles (1980)
underestimated the filling fraction because he included only northern HI
data, missing the most important Galactic quadrant 3, which is in the
southern sky and has more obvious superbubble dominance.  Theoretically,
that McKee (1993) neglects superbubble-shock-induced star formation in
which a superbubble shock triggers the formation of additional star
clusters. Heiles (1998) believes that Orion/Eridanus and Vela are a
result of this process, and this process might occur in the Magellanic
Clouds (Oey \& Smedley 1998; Points et al 1999).  Such correlated
cluster formation would produce ISM regions where correlated
superbubbles dominate---compensated for by other regions where they
don't (because the total rates are, presumably, known). 

\subsection{The amalgam of densely-packed superbubbles}

          A young superbubble is a clear, distinct entity.  It is
rejuvenated by successive SN, and these SN may be produced through
successive stellar generations.  The superbubble expands and, when the
sources of energy finally die out, slows down and merges with the
surroundings.  The walls remain atomic unless there are sources of
ionizing photons; portions of the walls that had been ionized begin to
cool and recombine when the sources disappear.  The interior HIM cools
adiabatically as it expands.  An expanding superbubble eventually runs
into volume occupied by its neighbors.  The walls fragment and become
the neutral interstellar clouds.  Often a different SN or superbubble
shock overtakes such clouds, which then end up inside the remnant bubble
as in the MO model and evaporate into the interior HIM (McKee \& Cowie
1977).

	The interstellar volume becomes an amalgam of all the gas
components. The HIM regions should connect and form tunnels as
envisioned by Cox \& Smith (1974).  The neutral gas should form walls of
these tunnels and perhaps be embedded within as clouds or fragments. WNM
that happens to be immediately adjacent to the HIM is partially ionized
by its X-rays, forming the MOWIM.  Some of the HI haphazardly turns to
RWIM when it is exposed to UV photons; after the ionizing photons turn
off,  the RWIM cools and recombines in a time-dependent, nonequilibrium
fashion. The walls contain the  occasional region where dense molecular
gas happens to have formed as a result of shocks, collisions, and random
turbulent intermittency (\S7).  

\section{The Two WIM's: RWIM and MOWIM \label{wimsection}}

	The WIM is revealed by observations of the diffusely-distributed
H$\alpha$ line (which provides the emission measure $EM = n_e^2L$) and
the dispersion of pulsars (which provides the dispersion measure $DM =
n_eL$).  There is an additional observational probe: interstellar 
scintillation/scattering of small diameter radio sources (pulsars and
extragalactic nuclei), which is parameterized by the scattering measure
SM.  The SM is not a simple quantity because it is a measure of
interstellar turbulence and small-scale variations in $n_e$.   

	The classical assumption is that the WIM consists of a single
type of gas---equivalently, that the {\it same} electrons produce the DM
and EM. With this assumption, Haffner et al (1999) use line intensity
ratios ${\rm NII} \over {\rm H}\alpha$ and   ${\rm SII} \over {\rm
H}\alpha$  to derive a typical WIM temperature $T \sim 8000$ K. Reynolds
et al (1998) use the intensity ratio  ${\rm OI} \over {\rm H}\alpha$  to
derive the fractional ionization, typically $\sim 0.9$ for $T = 8000$ K.
These approximate the conditions expected for ionization by the same
stellar photons that produce HII regions. This, together with the
filamentary morphology, leads us to the picture that much of the WIM
resides in filaments lying on superbubble walls, as in Eridanus (\S3.1).

	However, the data can be also be interpreted using a different
assumption, namely that {\it different} electrons produce the DM and EM.
These are the RWIM and the MOWIM. The RWIM is the HII-type WIM discussed
just above, and it produces most of the EM (that is, most of the
quantity $n_e^2L$ along any line of sight). The RWIM produces some or
most of the H$\alpha$ and nearly all the collisionally-excited heavy
element emission lines. The MOWIM is partially ionized WNM, and produces
most of the DM.  If the temperature of the MOWIM is low enough, then it
can produce significant H$\alpha$ even with its smaller EM, because the
H$\alpha$ emissivity $\propto T^{-0.9}$. 

\subsection{The need for two WIM components} \label{wimpressure}

	In this section we address the need for two WIM components on a
global scale. We do this by {\it assuming, temporarily,} that there is a
single WIM component and calculating its typical pressure. This pressure
is small compared to that of the CNM, and we argue that this is
unlikely. We then go on in the next two sections to discuss the two WIM
components.

\bigskip

\noindent{\it The single-component WIM pressure WITHOUT Interstellar
Scattering Data.} Reynolds (1991), in a classic paper, compared DM's of
distant high-latitude pulsars with EM's.  By assuming  that the WIM lies
in clouds of average electron density $n_a$ with fluctuations $\sigma$,
he derived

\begin{equation}
\label{reynoldseqn}
n_a = {EM/DM \over 1 + \sigma^2 / n_a^2}
\end{equation}

\noindent and used H$\alpha$-derived EM's and DM's
towards distant pulsars in globular clusters, together with the
assumption $\sigma^2 = 0$, to derive $n_a \approx 0.08$ cm$^{-3}$ at
$z=0$.  The WIM temperature is about 7000 K, a reliable value if it is
in thermal equilibrium and an upper limit if not, so the pressure is
${P \over k} = 2 n_e T \la 1100$ cm$^{-3}$ K. This is the typical
pressure of the WIM, and it is nearly four times smaller than that of
the CNM. 

\bigskip

\noindent{\it The single-component WIM pressure WITH Intersstellar
Scattering Data.}  Measurements of the Scattering Measure SM show that
the WIM's typical pressure is even smaller.  There are two types of
scattering, diffractive and refractive.  The former produces
randomly-distributed islands of intense emission in the frequency-time
plane, much like terrestrial atmospheric seeing, and the latter produces
effects like two-slit interference fringes---correlated islands, much
like elongated mountain chains such as the Tetons.  Examples of these
data are in Stinebring, Faison, \& McKinnon (1996) and Gupta, Rickett,
\& Lyne (1994).  These data allow the determination of the turbulence
spectrum in the WIM, which is typically very close to being Kolmogoroff.
 This is a power law with cutoffs at both small and large scales; the
large scale cutoff is called the outer scale. TC provide a summary. 

          Cordes et al (1991) interpret these data in a classic paper. 
The fluctuations that produce the observed scattering are tiny in size,
$\sim 10^5$ km, but there are good reasons to extrapolate the inferred
turbulence spectrum to larger size scales.  Moreover, the refraction
scattering is pronounced and is produced by larger (but still tiny: a
few AU) fluctuations; this indicates that the spectrum is even steeper
than Kolmogoroff.  The result is that the outer scale of the turbulence
spectrum is inferred to extend all the way up to the size of the clouds
and this means that, in equation~\ref{reynoldseqn}, one should take
${\sigma^2 \over n_a^2} \sim 1$.  This decreases $n_a$ still further, by
a factor 2 to about 0.04 cm$^{-3}$, so the typical pressure in the WIM
clumps becomes only ${P \over k} \la 600$ cm$^{-3}$ K.

\bigskip

\noindent{\it Commentary: WIM pressure and the two WIM's.} One might not
be surprised at a systematic difference in thermal pressure between the
WIM and the CNM. However, I think one should be surprised if the typical
WIM pressure were systematically {\it smaller} than the typical CNM
pressure, particularly by such a large factor.  We think of the
single-component WIM as portions of neutral shell walls, or fragments
thereof, that happen to have been exposed to ionizing stellar
photons---and when the neutral medium is ionized, its pressure increases
by at least a factor of two.  This idea is strengthened by the
observations of the filaments quoted in
\S3.1. 

	Let us suppose for the moment that the WIM pressure follows our
intuitive expectation and is significantly larger than the 600 cm$^{-3}$
K.  We derived the pressure using equation~\ref{reynoldseqn};
increasing the pressure requires changing the ratio ${EM \over DM}$. 
The only way to accomplish this is to assume that most of the observed
DM comes from a different electron component than the EM. To retain
approximate thermal pressure equality, this component must contribute
most of the observed DM and less of the observed EM, which pulls it
towards lower $n_e$; yet, in conflict, we want it to have a higher
pressure.  The only way out is to embed the electrons in one of the
primarily neutral phases.  The only reasonable such phase is the WNM,
because only it has small enough $n_{HI}$ to allow the needed $n_e$. 

\bigskip

\noindent { \it Commentary: the RWIM and the MOWIM.} Thus, we identify
the MOWIM with a partially ionized portion of the WNM.  This means that
portions of the WNM must have significant fractional ionization with
$X_e \ga 0.1$.  Our identification is not the first observational
indication of this association.  Indeed, Spitzer \& Fitzpatrick (1993;
SF93) have already made this suggestion on the basis of UV absorption
line data towards HD93521.  Moreover, the local cloud adjacent to the
Solar system has $X_e \sim 0.5$ (Redfield \& Linsky 2000; RL).  However,
these measurements are for individual regions; what's new here is our
claim that the MOWIM is {\it globally} important.

	Our argument suggests that the MOWIM provides most of the pulsar
DM's. It also suggests one observationally important item (Spangler
1991), which was brought to my attention by Cordes and Zweibel (private
communication). The degree to which the interstellar electrons produce
scattering fluctuates widely with position (Cordes et al 1991, TC). The
turbulence that produces the scintillation exists at such small
scales exists because of transfer of turbulent energy from large to
small length scales (Goldreich \& Sridhar 1995). The turbulence is
damped by ion-neutral collisions, whose importance increases to a
saturated value at small length scales. It is possible that the larger
neutral content in the MOWIM reduces the small-scale turbulence enough
to render it ineffective for interstellar scattering, leaving the RWIM
as the primary scattering medium.

\section{The MOWIM: Partially Ionized WNM \label{wnmionization}}

	In this section we discuss observations of the WNM and its
fractional ionization in specific regions. Some of these support the
idea of the MOWIM. 

\subsection{ The Dispersion Measure Imperative Assumption (DIMP)}

	In \S4.1 we introduced the MOWIM and suggested that portions of
the WNM are nontrivially ionized, contributing significantly to the
pulsar DMs.  Unfortunately, this idea is hard to test because the
optical/UV line data do not provide direct measurements of ionization
fraction for situations where $n_{HI}$ is small, as in the WNM. What
they {\it do} provide is direct measurements of $n_e$; the trick is to
somehow obtain $n_{HI}$ values to  associate with these $n_e$
measurements. We will accomplish this with the Dispersion Measure
Imperative Assumption. 

	Fitzpatrick \& Spitzer (1997, FS) is the culmination of an
excellent series of four optical/UV absorption line studies of
high-latitude gas.  The optical/UV data provide electron volume
densities $n_e$ at a sensitive level: for WNM components having
$N_{HI,20} \ga 0.1$, FS detect $n_e$ as small as $0.02$ cm$^{-3}$.  They
also detect one ionized component with $N_{e,20} \sim 1.3$ and $n_e \sim
0.2$ cm$^{-3}$.  These numbers suggest that FS can reliably detect {\it
all} the interstellar electrons over the ranges of volume and column
density that are relevant for DM and EM studies. 

	We believe the DM-producing electrons to be pervasive.  Thus,
they should exist on every sightline.  Because of the sensitivity of the
FS data to electrons, FS should detect the DM-producing electrons on
{\it every sightline}.  They don't find fully-ionized components, except
for one probably isolated small HII region.  Thus one is drawn toward
the {\it DM-Imperative Assumption}, or {\it DIMP}: 

	{\it We assume the WNM components to contain all the electrons
required for pulsar DM.}  This means that we assume each component to
have the global ratio of column density ${n_e \over n_{WNM}} =
{N_{e\perp} \over N_{WNM\perp}} \sim {0.5 \over 1.9}$ (from
Table~\ref{bigtable}); this corresponds to $X_e = 0.21$.  $N_{HI}$ and
$n_e$ are both known for some components, so we obtain $n_{HI}$---and
then the total thermal pressure. 

	Of course, this assumption is extreme. First, MO predict the
ionization fraction of the WNM to be highly variable: it depends on the
exposure to soft X-rays, which are atteunuated by small HI column
densities.  Second, it doesn't include the DM contribution of the
conventional, fully-ionized RWIM. 

\subsection{ Discussion of specific stars}

\noindent { \it The HD93521 sightline.} The HD93521 sightline (SF93)
contains no CNM (except one insignificantly small component with
$N_{HI,20} = 0.06$): the coldest gas component has upper limit $T \la
540$ K, and most of the other temperature upper limits are in the range
$T \la 4800 \rightarrow 6900$ K with derived $n_e = 0.055 \rightarrow
0.11$ cm$^{-3}$.  From the DIMP assumption we obtain typically $n_{HI}
\sim 0.2 \rightarrow 0.43$ cm$^{-3}$, yielding total thermal pressures
${P \over k} \sim 2000 \rightarrow 4000$ cm$^{-3}$ K.  

	Our quick discussion underscores the important conclusion of
SF93: along this sightline at least, {\it the DM-producing electrons
reside in the WNM}.  The only way to negate this conclusion is to reduce
$X_e$ from its assumed value of 0.21.  However, this raises $n_{HI}$ and
causes the thermal pressures to be much higher than a few thousand
cm$^{-3}$ K, which would make them exceed the upper limit for WNM in the
Wolfire et al (1995a; WHMTB) model. Slavin, McKee, \& Hollenbach (2000)
account very well for the HD93521 observations with soft X-ray photons
from old SN remnants. 

\bigskip

\noindent { \it The HD215733 sightline.} This sightline (FS) has only
one useful WNM component, which has $T = 730 \pm 70$ K (a value, not an
upper limit) and $n_e = 0.07$ cm$^{-3}$.  The DIMP assumption gives
$n_{HI} = 0.27$ cm$^{-3}$ and thermal pressure ${P \over k} = 300$
cm$^{-3}$ K.  The density is about ten times smaller than the
pressure-equilibrium value at $T = 730$ K. 

	If instead the thermal pressure is ${P \over k} = 3000$
cm$^{-3}$ K, then $n_{HI} \sim 4.1$ cm$^{-3}$, about twice the
equilibrium value for $T = 730$ K.  The ionization fraction is $X_e \sim
0.017$, about 5 times higher than equilibrium in the standard model of
WHMTB. 

	In terms of the MO paradigm of pressure equilibrium, the
latter possibility is much more reasonable.  But then we must ask,
``Where do the pulsar-DM electrons reside?'' Perhaps this sightline has
a very small DM?

\bigskip

\noindent { \it The HD149881 sightline.} In contrast to the others, this
sightline (Spitzer \& Fitzpatrick 1995; SF95) contains a truly ionized
component.  However, SF95 interpret it as a small HII region, probably
produced by HD149881 itself.  Such HII regions are presumably rare, so
the assumption that the DM-producing electrons are pervasive still
applies to the neutral components along this sightline.

	This star has three WNM components with $T \sim (6000, 840,
2000)$ K.  We quote their Na-determined electron densities, which we
regard as more reliable than the Ca values because the depletion of Na
is less variable.  The quoted electron densities are lower limits
because of the assumption that Na is undepleted.  SF95 provide values
for only the first and third components; they are $n_e \ga (0.049,
0.018)$ cm$^{-3}$. 

	The DIMP assumption gives $n_{HI} = (0.19, 0.069)$ and ${P \over
k} = (1700, 200)$ cm$^{-3}$ K.  The first component is in reasonable
agreement with thermal equilibrium.  The second component suffers
incompatibility with equilibrium to a similar extent as the HD325733
component. 

\bigskip

\noindent { \it The Colorado Model of the Local Interstellar Cloud.} The
local cloud just adjacent to the Solar system has well-determined
properties: $n_{HI} = 0.10$ cm$^{-3}$, fractional ionization $X_e =
0.52$ (exceeding the DIMP assumption!), and $T \sim 7000$ K.  These
combine to give the relatively high value ${P \over k} \sim 2200$
cm$^{-3}$ (RL). 

\subsection{Commentary on WNM ionization}

	The DIMP assumption works well for the first star, HD93521, and
also for the Local Interstellar Cloud.  It also works reasonably well
for the first component of HD149881.  We conclude that along some
sightlines, and in some WNM clouds, the DM-producing electrons reside
primarily in the WNM instead of the standard WIM. In other words, we
concude that the MOWIM really does exist and might be globally
important.

	In ionization equilibrium the total H-ionization rate is
$\zeta_{-15} = n_{HI} \left({X_e \over 0.05}\right)^2$ (KH, section
6.4), where $\zeta_{-15}$ is the total H-ionization rate in units of
$10^{-15}$ s$^{-1}$.  Taking $X_e = 0.2$ and $n_{HI}= 0.3$, these WNM
components have $\zeta_{-15} \sim 5$.  This seems very high, but the MO
model has it much higher, $\zeta_{-15} \sim 110$; it comes from soft XR
emitted by the adjacent HIM, and is attenuated very rapidly going inside
the cloud because of absorption by the HI. 

	This conclusion, that significant DM-producing electrons reside
in the WNM, is precisely opposite to that of section 6.4 of KH.  Their
conclusion was based on the assumptions of thermal pressure equality,
equilibrium for both temperature and ionization, and much smaller values
of $\zeta_{-15}$ based on observations of dense clouds.  The optical/UV
absorption line data make it clear that those assumptions are not
universally valid!

	The DIMP assumption works much less well for HD215733 and the
second component of HD149881 because it produces small values for HI
volume density and thermal pressure; we instinctively rebel against such
small values because they are so far from conventional ones.  Both these
components have $T \sim 800$ K.  This is an awkward temperature for
equilibrium models because it is in the thermally unstable region.  If
one prefers to discard the DIMP assumption for these components, then
the physical parameters are closer to conventional values.  However,
then one must ask where the DM-producing electrons reside on these
sightlines.  Perhaps the answer is that these sightlines don't contain
the expected electron column densities, with the possible ramification
that the DM-producing electrons are not very pervasive. 

	If, instead, these awkward components do contain a significant
fraction of the DM-producing electrons, then they (and presumably many
others) are greatly thermally underpressured with respect to their
surroundings.  This doesn't necessarily mean that the {\it total}
pressures in these structures are small, because generally speaking
macroscopic motions, magnetic field, and cosmic rays contribute most of
the interstellar pressure.  Perhaps the implied strong field strengths
explain the relatively large values of rms magnetic field strength
determined from studies of DM fluctuations (see reviews by Heiles 1995,
1996). 

\section{New Observations of the CNM and WNM \label{cnm-wnm}}

\subsection{ Temperatures from New 21-cm Line Observations
\label{radiotemps}}

	Tom Troland and I have been pursuing an intensive observational
program to measure properties of the CNM and WNM using 21-cm line
emission/absorption line observations at Arecibo Observatory, the
details of which will be forthcoming in future papers\footnote{I
emphasize that the results presented here are provisional and will
change in detail (but hopefully only in detail!).}.  Our interpretation
is new because we assume clouds of uniform temperature and  we perform a
physically self-consistent treatment of the radiative transfer, with
excellent results. 

	The absorption spectra consist of very obvious velocity
components and we represent their optical depths by Gaussians.  We
decompose the associated emission profile into two sets of Gaussians:
(1) a set with the same centers and widths as the absorption profile's
Gaussians; and (2) one or two additional Gaussians to make up the
difference.  The former set is the CNM and the latter the WNM. 
Interesting details include considering the fraction of WNM that lies in
front of the CNM and, also, the relative ordering along the line of
sight of the CNM components---nearby ones absorb the more distant ones. 
For most sources the CNM components plus one or two WNM
components produce a very good fit.  

\begin{figure} 
\plotone{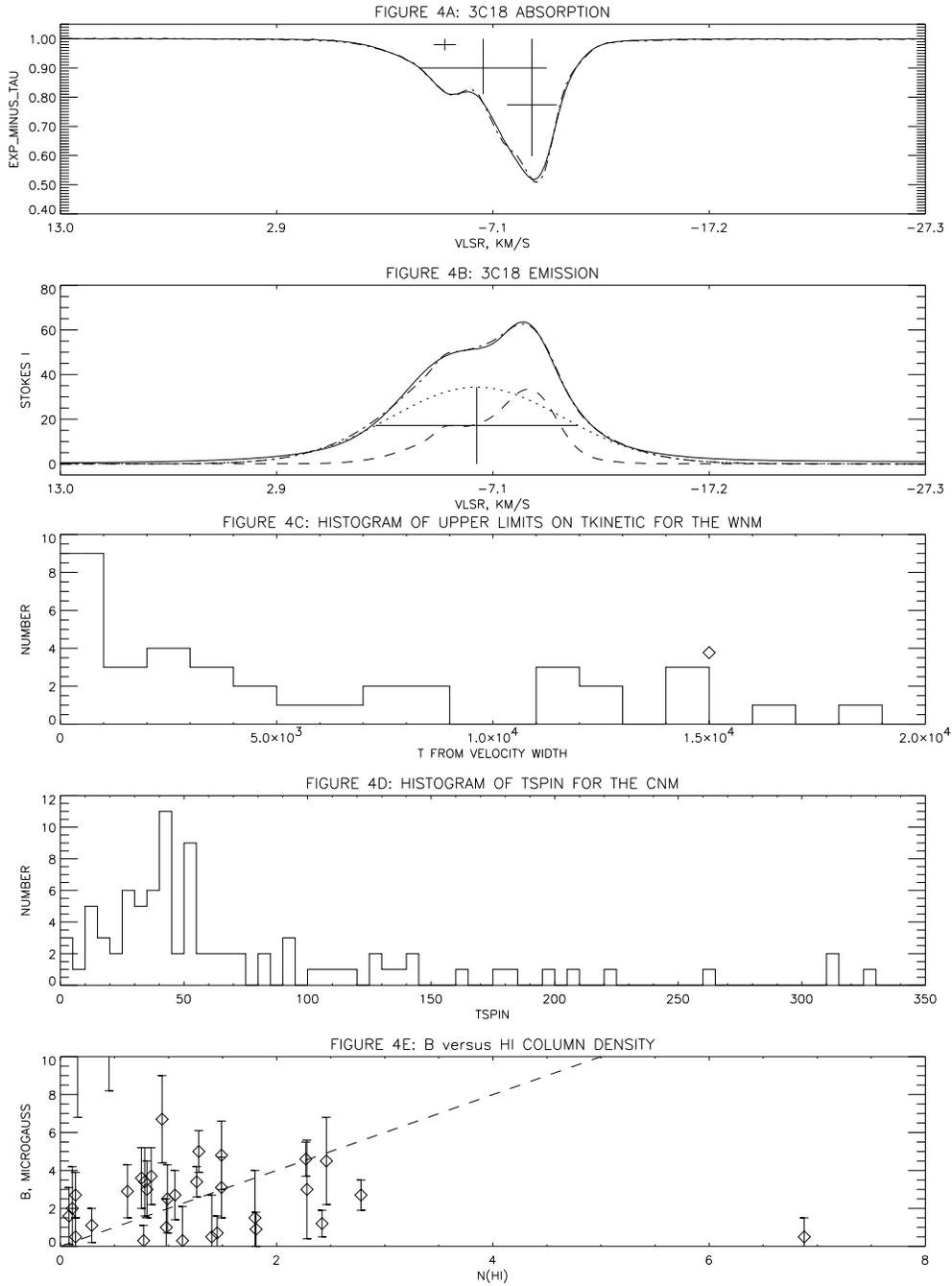} 
\caption{ {\bf A} and {\bf B} exhibit the 21-cm line absorption and
emission towards 3C18; the solid line is data and the dash-dot line the
fit.  Crosses indicate Gaussian component parameters.  In {\bf B} the
dashed line illustrates the contribution from the three CNM components
and the dotted line the WNM component.  {\bf C} is the histogram of
upper limits on kinetic temperature for the WNM components and {\bf D}
the histogram of spin temperatures of the CNM components.  {\bf E} plots
the observed line-of-sight magnetic field strength versus HI column
density (units: $10^{20}$ cm$^{-2}$) for the CNM components. 
\label{arecibo1}} \end{figure}

	We show the results for 3C18, which is one of the simplest
profiles and good for an illustrative example.  Figure~\ref{arecibo1}A
is the absorption spectrum.  The solid line is the observed absorption
spectrum, which we fit with the three CNM components whose depths and
halfwidths are indicated; the dash-dot line is the fit, which is almost
indistinguishable from the data.  Figure~\ref{arecibo1}B is the emission
spectrum.  The solid line is the observed profile, which we fit with
{\it (1)} the amplitudes of the three absorption Gaussians, keeping
their centers and widths fixed; plus {\it (2)} a single WNM component. 
The dashed curve is the contribution to the emission spectrum from the
absorption components and the dotted curve the contribution from the WNM
component, which is unabsorbed by the CNM because we take all the WNM as
lying in front the CNM, which placement provides the lowest residuals. 
The full fitted curve is the sum, shown as dash-dot, which is a good
fit. 

	The WNM component has halfwidth 10.0 km s$^{-1}$, which
corresponds to purely thermal broadening at $T = 2200$ K; this is an
upper limit on the kinetic temperature. For the three CNM components,
left-to-right on Figure~\ref{arecibo1}A, the spin temperatures  are $46
\pm 9$, $43 \pm 6$, and $32 \pm 1$ K; their Zeeman splittings (data not
shown) give line-of-sight field strengths $0.3 \pm 12.1$, $10.7 \pm
5.8$, and $-3.6 \pm 1.6$ $\mu$G. 

	The WNM component contributes $N_{HI,20} = 3.2$ and the three
CNM components $N_{HI,20} = 1.8$, where $N_{H,20}$ is the HI column
density in units of $10^{20}$ cm$^{-2}$.  The WNM/CNM ratio is
$\sim 1.8$, significantly higher than the average of $\sim 1$. About
half the HI is WNM, much higher than the MO predicted fraction of $\sim
0.04$. 

\subsection{ Temperatures from optical/UV absorption line
observations \label{uvtemps}}

	For kinematic studies, optical/UV absorption line studies  of
high-latitude gas have the advantage that the thermal broadening of
heavy elements is small, so closely-spaced velocity components are more
distinctly revealed than with the 21-cm line.  Their disadvantage is
that derived temperatures aren't very accurate because they are obtained
by comparing heavy-element and HI linewidths; the HI line comes from a
much larger angular area so the nonthermal component of its width may be
larger than that of the heavy element lines, raising the derived
temperatures above the actual ones.  In particular, optical/UV
temperatures for CNM gas are not at all accurate. 

	One principal result of FS is that each Gaussian that one would
fit to a 21-cm emission line consists, in the optical/UV absorption line
data (and thus reality), of a superposition of several closely-spaced
components.  The relation between the optical/UV lines and the 21-cm
emission line is much like that between the 21-cm absorption and
emission lines for 3C18 in Figure~\ref{arecibo1}A: the separations and
widths are not dissimilar.  It is natural, then, to assume that the
21-cm absorption line data represent the complete picture of CNM gas. 
It would be nice to confirm this by using exactly the same sightlines
for the radio and optical measurements. 

\subsection{ TEMPERATURE OF THE WNM \label{wnmtemp}}

\noindent { \it The new 21-cm line data.} The WNM Gaussian widths
provide upper limits for kinetic temperature $T_K$. 
Figure~\ref{arecibo1}C exhibits a histogram of these limits.  About half
the components have $T_K < 5000$ K; because these components are not
visible in absorption, their spin temperatures exceed $\sim 500$ K. 
This range, $500 \rightarrow 5000$ K, is the thermally unstable range
that separates CNM from WNM.  

	Our Arecibo data show this departure from thermal stability in a
statistically convincing manner.  Previous studies have hinted at this
result.  In emission/absorption studies, Mebold et al (1982) decomposed
emission line profiles into Gaussians, with similar results; however,
they didn't explicitly point out this departure.  Verschuur \& Magnani
(1994) and Heiles (1989) analyzed emission profiles and found pervasive
components with widths in this range, but without absorption data could
not conclusively state that the kinetic temperatures were indeed so
high. 

\bigskip

\noindent { \it The FS and Colorado data.} FS derive WNM temperatures by
comparing heavy-element and HI linewidths; we restrict our summary to
those WNM components having $N_{HI,20} > 0.1$ that also have
well-determined temperatures or upper limits.  The character of their
results depends on the particular sightline. 

	For HD93521 [$(\ell, b) = 183^\circ, 62^\circ)$] they found four
components, all with high upper limits on kinetic temperature $\la 4900
\rightarrow 6900$ K.  For HD149881 [$(\ell, b) = 31^\circ, 36^\circ)$]
they found two components with kinetic temperatures $840 \pm 150$ and
$2000 \pm 400$ K.  For HD215733 [$(\ell, b) = 85^\circ, -36^\circ)$]
they found two components with kinetic temperatures $730 \pm 70$ and
$2500 \pm 280$ K. Finally, the Local Interstellar Cloud has $T \sim
7000$ K and ${P \over k} \sim 2200$ cm$^{-3}$ (RL).

\bigskip

\noindent{\it Commentary on WNM temperature } We conclude that many WNM
components are in the thermally unstable temperature range.  Pressures
in the WNM apparently have wide variations: the Local Cloud and the
sightline towards HD93520 are characterized by ${P \over k}$ in the
thousands, while the other two have much lower temperatures and might
have ${P \over k}$ in the hundreds if the DIMP assumption of \S5.1
applies. Thermally unstable temperatures should not often occur in the
MO model, and such low pressures should be rare.

	The isobaric stability criteria don't necessarily apply in an
obvious way to the WNM.  WNM structures have low densities, $n_{HI} \sim
0.1 \rightarrow 0.4$ cm$^{-3}$ for the FS's five WNM components towards
HD93521 (\S5.2), and should occupy relatively long lengths along the
sightline, tens of pc, so the sound crossing time should be comfortably
longer than the thermal cooling time.  

	WHMTB find Ly$\alpha$ cooling to dominate the WNM, even at the
low fractional ionizations that occur in equilibrium.  If the ionization
fractions are as large as required for the DIMP assumption in \S5.1 to
be valid, meaning that the DM-producing electrons do indeed occur in the
WNM, then some H$\alpha$ lines in the WHAM survey should have parallels
in the 21-cm line emission from low-temperature WNM/MOWIM. 

\subsection{ Temperature of the CNM} 

	For the CNM, spin temperature is the same as kinetic
temperature.  Figure~\ref{arecibo1}D exhibits the histogram of derived
spin temperatures for all CNM components.  Most components have $T_S <
75$ K.  This is in marked contrast to previous results, where histograms
had broad peaks over the ranges $20 \rightarrow 140$ K (Mebold et al
1982) and $50 \rightarrow 300$ K (Dickey, Salpeter, \& Terzian 1978,
DST; Payne, Salpeter, \& Terzian 1982, PST).  Our range is narrower and
there is a suggestion that the histogram is doubly peaked, with one peak
in the range $10 \rightarrow 20$ K and the other $30 \rightarrow 75$ K. 

	Consider first the higher temperature $30 \rightarrow 75$ K
range.  This agrees very well with theory. WHMTB included all known
processes in calculating their standard model, for which the CNM
equilibrium temperatures range from $25 \rightarrow 200$ K (the
corresponding densities are $n_{HI} \ga 1000 \rightarrow 4$ cm$^{-3}$). 
Our observed temperature range is smaller and corresponds to $n_{HI} \ga
250 \rightarrow 20$ cm$^{-3}$ and ${P \over k} = 10000 \rightarrow 1500$
cm$^{-3}$ K.  These numbers are in accord with current ideas about the
CNM.  

	Modern theoretical models do not predict our low-temperature $10
\rightarrow 20$ K range.  However, the theoretical interpretation of
such low temperatures is actually quite straightforward.  In the days
when Spitzer (1978) wrote his famous textbook, heating by photoelectric
emission from dust grains was not a well-accepted process and Spitzer
calculated the CNM equilibrium temperature assuming that only the
classical mechanisms prevailed, namely heating by photoionization of
Carbon and cooling by electron recombination onto ionized Carbon.  The
equilibrium temperature is 16 K and is independent of the abundance of
Carbon and of the starlight intensity.  This is precisely in the middle
of our low-temperature range!

	In order that CNM gas {\it not} be heated by photoelectric
emission from grains, the gas cannot contain grains---more specifically,
the gas cannot contain the particular kinds of grain that heat the gas. 
In WHMTB's theory, these are small grains ($\la 300$ \AA) and PAH's.  It
seems difficult to produce gas without such grains.  One cannot rely on
strong shocks, because they add to the small-grain population by
shattering large grains (Jones, Tielens, \& Hollenbach 1996).  It seems
that one must destroy all the grains, perhaps by cycling the gas through
the HIM phase, or let the CNM cloud sit quiescentally for a long time so
that small grains coagulate into large ones. 

	These parcels of very cold gas shouldn't contain small grains
and PAH's! This prediction can be checked observationally from far-IR
emission maps, and perhaps from Ca absorption lines (gaseous Ca is
usually rare in the ISM because of depletion onto dust grains). Such
cold gas was invoked by Heiles (1997) to help understand the existence
of tiny-scale atomic structure; the observed existence of such cold gas
is encouraging for that interpretation. 

\bigskip

\noindent{\it The Morphology of the CNM.} Typical Gaussian components
of the CNM contain $N_{HI,20} \sim 1$.  Consider a CNM cloud having $n_H
= 100$ cm$^{-3}$.  If it is spherical, then its diameter is 0.3 pc.  The
scale height of the CNM is $\sim 100$ pc, so our CNM cloud is typically
at $50 \over \sin (b)$ pc distance and occupies angle $\sim {20 \over
\sin (b)}$ arcmin.  This is uncomfortably close to the Arecibo beam
diameter.  Nevertheless, DST and PST found that most of the emission
fluctuations occur on larger angular scales, and our Gaussian
decompositions show that roughly half of the total emission comes from
the CNM.  Moreover, FS find good correspondence between optical/UV
components along the sightline to a star and 21-cm line emission
components observed with a 20 arcmin beamwidth.  This is somewhat of a
conflict. 

	It makes sense to drop the idea of spherical clouds;  there is
much evidence for nonspherical structures in the ISM.  To resolve the
conflict, CNM structures need to be relatively large across the line of
sight compared to along the line of sight.  This can be achieved if
their morphologies are sheetlike.  This makes sense from another aspect:
the CNM is supposed to be formed by shocks. 

\bigskip

\noindent{\it Magnetic field strengths in the CNM.} Even for
``neutral'' atomic gas there is plenty of residual ionization to satisfy
flux freezing over many tens of Myr.  Accordingly, the magnetic field
strength should be large in the CNM: both the gas and also the magnetic
field have been swept up from within the entire volume of the
superbubble.  This idea is confirmed by observed large field strengths,
$\sim 10$ $\mu$G, which are observed in gas that resides in
morphologically obvious superbubble walls (Heiles 1989; Myers et al
1995). 

	Our current Zeeman splitting measurements apply not to obvious
supershell walls, but rather to sightlines towards radio sources and are
thus randomly selected.  We measure very low field strengths or upper
limits, typically no more than a few $\mu$G.  Figure~\ref{arecibo1}E
plots observed field strength versus $N_{HI,20}$ for these CNM Gaussian
components, together with a dashed line that represents the approximate
boundary between magnetism dominating gravity for self-gravitating
clouds (which may be completely irrelevant: gravity plays no role in
these clouds).  All points having $N_{HI,20} > 1.5$ fall below the line.
Moreover, most of the reliably measured points have $B \la 4$ $\mu$G, in
marked contrast to the $\sim 10$ $\mu$G field strengths in
morphologically obvious supershell walls.  ``Morphologically obvious''
means that the walls are seen edge-on; in contrast, the random lines of
sight preferentially sample sheets (\S6.4) that are {\it not} edge on.

	The new CNM field strengths are not higher than the
volume-average field strength of about 5 $\mu$G derived from various
observations independently by Heiles (1996) and Ferri\`ere (1998).  The
volume densities that characterize this volume-averaged field are very
small, much smaller than the $n_{HI} \ga 100$ cm$^{-3}$ of our CNM
clouds.  With flux freezing, field strength should increase with volume
density.  Thus flux freezing doesn't apply to these randomly-chosen CNM
clouds.  Our result, that the CNM does not necessarily contain high
magnetic fields, is consistent with the analysis of other observations
by Padoan \& Nordlund (1999). 

\section{ The Turbulent ISM \label{turbulence}}

	MO dealt with turbulence by considering how interstellar clouds
as discrete entities would be buffeted by passing SN shocks, much as
ocean waves toss jellyfish about. A number of recent numerical
experiments emphasize a different type of turbulence which has no
discrete entities, but instead has continuous distributions of physical
parameters, much as waves beating on a rocky headland produce relative
fluid motions with localized, temporary regions of high vorticity or
velocity dispersion; cloud boundaries are arbitrary surfaces that
dissipate with time  (e.g.~Ballesteros-Paredes, V\'azquez-Semadeni, \&
Scalo 1999; Smith, Mac Low, \& Heitsch 2000; Padoan \& Nordlund 1999; 
V\'azquez-Semadeni, Gazol, \& Scalo 2000).   The state of the CNM and
WNM is not completely controlled by the standard phase diagram, but also
 by the vissicitudes of turbulent flow. The numerical experiments
include most relevant physics, including magnetic fields, but are
two-dimensional.

	I find the numerical experiments attractive.  Observations make
it clear that turbulence is important in the ISM on a wide range of
scales, from tiny (interstellar scattering, \S4.1) to huge
(supershells). To the eye, the pictorial results of the  numerical
experiments actually look like the data---an excellent beginning, but
insufficient for quantitative comparison. This is quite a contrast from
the MO concept of shocks overtaking self-contained clouds whose mass
stays inside the boundaries.

	Extreme departures from average conditions are known as
intermittency. To my knowledge, this was introduced to astronomy in an
observational context as broad, nonthermal line wings by Falgarone \&
Phillips (1990) and in interpretive and theoretical contexts by
Falgarone \& Puget (1995) and Falgarone, Pineau Des Forets, \& Roueff
(1995).  Intermittency produces highly variable, non-Gaussian
distributions in space and time of velocity, vorticity, temperature, and
other physical variables.  The short-lived extreme conditions can
produce observed nonthermal motions, can decouple grains and gas, and
can initiate chemical reactions. 

	These numerical experiments have only recently begun and it is
too soon to say that they really do describe the observations better
than the MO paradigm. We look forward to further developments in this
field because it offers a new vista in comparing with observations, and
the comparisons seem to be favorable. 

\section{ The McKee/Ostriker Model: Paradigm?}  

	At the first Tetons meeting, KH reviewed the observations in
terms of the MO model with great success. How do things stand now? Here
is my personal view, not necessarily shared by others:

	$\bullet$ {\bf HIM:} MO works well---a plus (\S2.1).

	$\bullet$ {\bf RWIM:} Although MO doesn't predict this, others
do and this is not necessarily a negative for MO (\S2.2, 4).

	$\bullet$ {\bf MOWIM:} We suggest that this partially-ionized
WNM, which produces pulsar DM but not much else including interstellar
scattering, exists globally. This is contrary to most current
observational thought. MO predicts this component---a big plus (\S2.3,
4.1).

	$\bullet$ {\bf WNM mass:} MO predict this to be a minor part of the
ISM. Observations show that its mass is equal to, perhaps more than, the
CNM---a minus (\S6.1). 

	$\bullet$ {\bf WNM temperatures:} Our new 21-cm line data show
that significant portions of WNM are below 5000 K and thermally
unstable---a big minus (\S6.3).

	$\bullet$ {\bf CNM temperatures:} The new CNM temperatures agree
well with MO---a plus (\S6.4).

	$\bullet$ {\bf CNM morphology:} We argue that the CNM is
sheetlike, not spherical as envisioned by MO---a minus (\S6.4).

	$\bullet$ {\bf Magnetic fields and cosmic rays:} MO don't
include these in their theory. Their pressures are high and presumably
they have a big effect (\S1).

	$\bullet$ {\bf Turbulence:} It is not clear to me how turbulence
relates to MO (\S7).

	$\bullet$ {\bf Net Score:} One big plus and one big minus. Two
plusses and two minuses. Some questions on things not included. Of
course, high-quality data are sparse and much of our discussion is based
on unpublished data. Does all this make the MO model a paradigm? 

\acknowledgments

	It is a pleasure to acknowledge  conversations with John Dickey,
Alex Lazarian,  Pablo Padoan, and Enrique V\'azquez-Semadeni.
Discussions with Jim Cordes, Jeff Linsky, and Ellen Zweibel were
especially helpful and stimulating, resulting in a great improvement on
the first draft of this paper. The prize goes to Chris McKee, who
responded to the first draft with a huge set of comments that not only
reversed some of my misconceptions regarding MO but also stimulated new
ideas; however, this certainly does not mean that my understanding of MO
is complete or that he endorses all opinions expressed herein!

          This work was partly supported by NSF grant AST9530590 to the
author.

\end{document}